\documentclass[journal]{IEEEtran}

\ifCLASSINFOpdf
\else
\fi
\usepackage{amssymb}
\usepackage{amsmath}

\newtheorem{teiri}{Theorem}
\newtheorem{prop}{Proposition}

\newtheorem{rei}{Example}
\newtheorem{hodai}{Lemma}

\newtheorem{teigi}{Definition}
\newtheorem{remark}{Remark}

\begin{document}

\title{A Generalization of Nonparametric Estimation and On-Line Prediction for Stationary Ergodic Sources}

\author{Joe~Suzuki,~\IEEEmembership{Member,~IEEE,}
\thanks{J. Suzuki is with Osaka University.}
}

%
%

\markboth{}%
{Shell \MakeLowercase{\textit{et al.}}: Bare Demo of IEEEtran.cls for Journals}
%



\maketitle

\begin{abstract}
We propose a learning algorithm for nonparametric estimation and
on-line prediction for general stationary ergodic sources.
The idea is to prapare many histograms and estimate
the probability distribution of the bins in each histogarm.
We do not know a priori which histogram expresses the true distribution:
if the histogram is too sharp, the estimation captures the noise too
much (overestimation).
To this end, we weight those  distributions to obtain the estimation of 
the true distribution. As long as the weights are positive, we obtain a desired
property: the Kullback-Leiber information divided by the number $n$ of
examples diminishes as $n$ grows.
\end{abstract}



%
\IEEEpeerreviewmaketitle


\section{Introduction}

In machine learning, the problems of estimating a probabilistic rule given examples and  
predicting next data from the past sequence have been intensively considered for many years.
However, it seems that most of the effort has been devoted to the case assuming that 
each example consists of attributes taking finite values.
This paper deals with nonparametric estimation and on-line prediction 
assuming that the data sequence available for learning have been emitted by a general stationary
ergodic process. 

In this paper, by nonparametric estimation, we mean to captures the stochastic process that has emitted a given sequence without assuming 
that each random variable takes a finite value. If the random variables take finite values, then we only need to relative frequencies 
or its modifications to estimate the conditional probability $P(x_j|x_1,\cdots,x_{j-1})$ of next data $x_j$
 given the past sequence $x_1,\cdots,x_{j-1}$, so that
 the whole distribution 
 $$P(x_1,\cdots,x_n)=\prod_{j=1}^n P(x_j|x_1,\cdots,x_{j-1})$$ 
 can be estimated exactly as the sequence length $n$ grows, which is assured by the law of large numbers.
However, we are not sure what to do in general situations in which $\{X_j\}_{j=1}^\infty$ may not take finite values.
A similar situation will happen in on-line prediction that needs to estimate $P(x_j|x_1,\cdots,x_{j-1})$ as well.

To this end, we take an information theoretical approach.
Suppose we wish to construct an algorithm to 
compress data sequences as short as possible 
(the compressed data should be uncompressed to the original via another prescribed algorthm).
The best way would be to utilize the probability $P$ of the whole sequences to encode each sequence based on the framework 
of information theory \cite{gallager,cover}.
However, if such  knowledge is not available, we need to estimate $P$ from one given sequence.
If the process obeys the law of large numbers, say for stationary ergodic sources, the whole relative frequencies converges to definite values
 so that we can construct an estimation $Q$ of the original unknown $P$.
Then, we can encode the given one sequence $x_1,\cdots,x_n$  length roughly 
$$-\log Q^n(x_1,\cdots,x_n)$$
based on the $Q$.
Thus, the goal is to minimize the difference $D(P^n||Q^n)$ if the expectation 
$$E[-\log Q^n(X_1,\cdots,X_n)]$$
 should be short,
where $D(\cdot||\cdot)$ is  the Kullback-Leibler information \cite{kullback} 
which is useful for evaluating estimation errors.
Recent information theory covers the case not assuming the a prior knowledge of the stochastic nature of the sequences to be encoded  (universal coding).
We call a stochastic process simply  a source as used in information theory \cite{gallager}.

One might think why only the minimization of\\ $D(P^n||Q^n)$ is considered as a criterion of estimation and prediction
while many other ways to evaluate the performaces are available, for example,
$$e_n:=|P^n(x_n|x_1,\cdots,x_{n-1})-Q^n(x_n|x_1,\cdots,x_{n-1})|$$
may be a better alternative to $D(\cdot||\cdot)$. However, it is known \cite{gyorfi}\cite{ryabko2} that for any estimation $Q^n$, there exists 
a stationary ergodic source $P^n$ such that $\limsup_{n\rightarrow \infty}e_n>0$ with probability one.

Now let us we focus on the continuous data.
Recently, due to Boris Ryabko \cite{ryabko},  a great progress has been made  on the problems of 
 nonparametric estimation and on-line prediction for continuous data.
The idea is to estimate the probability density function for the source $\{X_j\}_{j=1}^\infty$.
Let $\{A_i\}_{i=0}^\infty$ be an increasing sequence of finite partitions of $\mathbb R$.
Then, we can estimate the probability $P_i^n$ of  the concatinations of $A_i$ of length $n$ to construct the estimation
$Q_i^n$ for each $i$. If we divide the original $P_i^n$ and estimation $Q^n_i$ by the volume of dimension $n$, then we can obtain
the probability density function $f_i^n$ and its estimation $g_i^n$  for each $i$, respectively.
Then, if we mixture the estimations $g_i^n$ by nonzero weights $\{\omega_i\}$ such that $\sum_i \omega_i=1$ and $\omega_i>0$,
then $g^n:=\sum_i\omega_i g_i^n$ can be an estimation of the probability density function $f^n$.
Ryabko proved that such an estimation $g^n$ has a disirable properties for nonparametric estimation and 
on-line prediction such as $D(f^n||g^n)/n\rightarrow 0$ ($n\rightarrow \infty$).

However, several problems remain even in Ryabko's inspiring framework:
\begin{enumerate}
\item The  probability density function $f$ should exist.
\item The differential entropy  of $f_i$ should converge to that of $f$ as $i\rightarrow \infty$.
\end{enumerate}

\begin{rei}\rm \label{rei2}
Suppose that random variable $X_j$ independently obeyes the distribution function
\begin{eqnarray*}
F_{X_j}(x)=
\left\{
\begin{array}{ll}
0&x<-1\\
\frac{1}{2}&-1\leq x<0\\
\int_{0}^x \frac{1}{2}g(t)dt&0\leq x
\end{array}
\right.
\end{eqnarray*}
with $\int_{0}^\infty g(x)dx=1$.
Then, $X_j$ does neither take  a finite value, nor have $f_{X_j}$ such that $$F_{X_j}(x)=\int_{-\infty}^x f_{X_j}(t)dt$$
for $x\in {\mathbb R}$ although $\{X_j\}_{j=1}^\infty$  is stationary ergodic.
\end{rei}
The proposed method generalizes the finite and continous cases as special cases.

For the proposed learning algorithm for nonparametric estimation and on-line prediction,
we do not have to know a priori whether the source is discrete or continuous, whether
the source has a probability density function when it is continuous. We only need to 
know the source is stationary ergodic, and the same algorithm can be applied to given data if the source is either discrete or continuous.
The derivation is very simple: unlike the original paper by Ryabko \cite{ryabko},
this paper applies A. Barron's generalized Shannon-MacMillian-Breiman theorem to obtain the general result.

Section 2 illustrates the main result using a typical example.
Sections 3 and 4 give  existing results for finite and continuous sources, respectively.
In particular, Section 4 summarizes Ryabko's original results on 
nonparametric estimation and on-line prediction without proof.
Those results in the two sections are extended to the general case in Sections 5.
Section 6 gives how the main result works in several cases
including when the existing methods cannot deal with.
In Sections 7, the main result is applied to on-line prediction. 
Section 8 concludes this paper with future topics.

\section{The idea: Histogram Weighting}

We first illustrate the main result using a typical example.
Suppose $\{X_k\}_{k=1}^n$ is Independent and identically-distributed and takes values in
$X_i(\Omega)\subseteq [0,1)$.
We wish to estimate the density function $f$ from examples $X_k=x_k\in [0,1)$, $k=1,\cdots,n$.
To this end, we prepare several histograms as follows.

\begin{tabular}{ll}
Level $0$:& $A_0=\{[0,1/2), [1/2,1)\}$ consisting of two bins\\
Level $1$:& $A_1=\{[0,1/4), [1/4,1/2),[1/2,3/4),[3/4,1)\}$\\& consisting of four bins\\
$\ldots$&$\ldots$\\
Level $i$:& $A_i$ consisting of $2^{i+1}$ bins\\
$\ldots$&$\ldots$\\
\end{tabular}

For each level $i=0,1,\cdots$, we can estimate the discrete distribution by counting the frequencies
of $2^{i+1}$ bins to obtain the estimate $Q_{n,i}(a)$ of the bin probabilities $P_i(a)$ for $a\in A_i$, $i=0,1,2\cdots$.
We do not know a priori which histogram $Q_{n,i}$ is the closest among $i=0,1,\cdots$ to the true distribution $P$.
Then, we approximate the density function $f(x)$ by $g_n(x)=\sum_{i=0}^\infty \omega_ig_{n,i}(x)$ with histograms
$\displaystyle g_{n,i}(x):=\frac{Q_{n,i}(s_i(x))}{2^{i+1}}$ and weights $\{\omega_i\}_{i=0}^\infty$ such that $\omega_i>0$ and $\sum_{i=0}^\infty \omega_i=1$,
where $s_i: [0,1)\rightarrow {A_i} $ is the projection.

Ryabko proved that the Kullback-Leibler information between $f$ and $g_{n}$ converges to zero as 
$n\rightarrow \infty$. 
However,  what if the density function $f$ does not exist as in Example \ref{rei2} ?
That is the problem we address in this paper.


\section{Finite Sources}\label{sec2}

Let $\{X_n\}_{n=1}^\infty$ be a stationary ergodic source expressed by probability $P^\infty$
generating each $X_n$ in a finite set $A$.
The entropy of the source is defined by \cite{gallager}
$$H(P^\infty):=\lim_{n\rightarrow \infty}-\frac{1}{n}\sum_{x^n\in A^n}P^{n}(x^n)\log P^{n}(x^n)$$
(such a limit always exists for stationary sources).

By coding, we mean any mapping
 $\varphi^n: A^n\rightarrow \{0,1\}^*$
 satisfying
$$\varphi^n(x^n)=\varphi^n(y^n) \Longrightarrow x^n=y^n$$
for $x^n,y^n\in A^n$.
It is known that 
$$\sum_{x^n\in A^n}2^{-|\varphi^n(x^n)|}\leq 1$$ (Kraft's inequality) for such $\varphi^n$, and that
if $l^n: A^n\rightarrow {\mathbb N}$ satisfies  
$$\sum_{x^n\in A^n}2^{-l^n(x^n)}\leq 1\ ,$$
there exists a coding $\varphi^n$ such that $|\varphi^n(x^n)|=l^n(x^n)$ for $x^n\in A^n$,
where we denote $|z|=m$ when $z\in \{0,1\}^m$. 
We can construct $\varphi^n$ from $P^n$ such that  \cite{gallager}
$$\lim_{n\rightarrow \infty}E\frac{|\varphi^n(X_1,\cdots,X_n)|}{n}=H(P^\infty)\ .$$

Even without knowledge of $P^n$, we can construct a coding
$\varphi_*^n$ satisfying
\begin{equation}\label{eq101}
\lim_{n\rightarrow \infty}\frac{|\varphi_*^n(x_1,\cdots,x_n)|}{n}=H(P^\infty)
\end{equation}
with probability one and
\begin{equation}\label{eq102}
\lim_{n\rightarrow \infty}E\frac{|\varphi_*^n(X_1,\cdots,X_n)|}{n}=H(P^\infty)
\end{equation}
for all stationary ergodic $P^\infty$ (universal coding) \cite{cover}.
\begin{hodai}[Shannon-MacMillan-Breiman \cite{gallager}]\rm \label{L1}
$$\lim_{n\rightarrow \infty}\frac{-\log P^n(x_1,\cdots,x_n)}{n}=H(P^{\infty})$$
with probability one
for all stationary ergodic $P^\infty$.
\end{hodai}

Let
$$Q^n(x_1,\cdots,x_n):=
{2^{-|\varphi_*^n(x_1,\cdots,x_n)|}}
\ ,$$
for $x_1,\cdots,x_n\in A$. 
From (\ref{eq101})(\ref{eq102}) and Lemma \ref{L1}, we have the following proposition.
\begin{prop}[Ryabko \cite{ryabko2}]\rm  \label{prop1}
\begin{enumerate}
\item 
\begin{equation}\label{eq301}
\frac{1}{n}\log \frac{P^n(x_1,\cdots,x_n)}{Q^n(x_1,\cdots,x_n)}\rightarrow 0
\end{equation}
with probability 1, and
\item 
\begin{equation}\label{eq302}
\frac{1}{n}\sum_{x_1,\cdots,x_n\in A}P^n(x_1,\cdots,x_n)
\log \frac{P^n(x_1,\cdots,x_n)}{Q^n(x_1,\cdots,x_n)}\rightarrow 0
\end{equation}
\end{enumerate}
for all stationary ergodic sources $P^\infty$ 
as $n\rightarrow \infty$.
\end{prop}

\section{Continuous Sources}

This section summarizes Boris Ryabko's pioneering work on continuous sources with a probability density function.

Let $\{X_n\}_{n=1}^\infty$ be a stationary ergodic source expressed by probability density function  $f^\infty$
generating each $X_n$.
The differential entropy of the source $f^\infty$ is defined by
$$h(f^\infty):=\lim_{n\rightarrow \infty}-\frac{1}{n}\int f^{n}(x^n)\log f^{n}(x^n)$$
(such a limit always exists for stationary sources).


Let $\{A_i\}_{i=0}^\infty$ be an increasing sequence of finite partitions of $\mathbb R$
that asymptotically generates the Borel $\sigma$-field $\cal B$, so that each element in $A_i$
is a non-empty disjoint subset of $\mathbb R$.
Let $s_i:{\mathbb R}\rightarrow A_i$ be the projection $x\mapsto a\ni x$.
For each $i=0,1,\cdots$, since $P_i^\infty$ is stationary ergodic over the finite alphabet $A_i$, 
we can construct a universal coding $\varphi_i^n$ for $A_i^n$, and define
$$Q_i^n(a_1,\cdots,a_n):=
{2^{-|\varphi_i^n(a_1,\cdots,a_n)|}}
\ ,$$
for $a_1,\cdots,a_n\in A_i$. For each $i=0,1,\cdots$, we define
$$f_i^n(x_1,\cdots,x_n):=\frac{P_i^n(s_i(x_1),\cdots,s_i(x_n))}{\lambda_i^n(s_i(x_1),\cdots,s_i(x_n))}$$
and
$$g_i^n(x_1,\cdots,x_n):=\frac{Q_i^n(s_i(x_1),\cdots,s_i(x_n)}{\lambda_i^n(s_i(x_1),\cdots,s_i(x_n))}$$
for $(x_1,\cdots,x_n)\in {\mathbb R}^n$,
where $\lambda_i^n(a_1,\cdots,a_n)$ is the Lebesgue measure of $(a_1,\cdots,a_n)\in A^n_i$.
Let $\{\omega_i\}_{i=0}^\infty$ be such that
$\sum_{i=0}^\infty \omega_i=1$ and $\omega_i>0$.
Then, we can define the density estimation $g^n$ as follows: for $(x_1,\cdots,x_n)\in {\mathbb R}^n$,
\begin{equation}\label{eq02}
g^n(x_1,\cdots,x_n):=\sum_{i=0}^\infty \omega_i g_i^n(x_1,\cdots,x_n)
\end{equation}

\begin{hodai}[Shannon-MacMillan-Breiman \cite{barron}]\rm \label{L3}
\begin{equation}\label{eq201}
\lim_{n\rightarrow \infty}\frac{-\log f^n(x_1,\cdots,x_n)}{n}=h(f^{\infty})
\end{equation}
with probability one for all stationary ergodic $f^\infty$.
\end{hodai}
We also consider the differential entropy $h(f_i^\infty)$ of the stationary ergodic source $f_i^\infty$ for each $i=0,1,\cdots$
If we apply Lemma \ref{L3} to the probability density function $f_i^\infty$, we have 
\begin{equation}\label{eq202}
\lim_{n\rightarrow \infty}\frac{-\log f_i^n(x_1,\cdots,x_n)}{n}=h(f_i^{\infty})
\end{equation}
with probability one.

\begin{prop}[Ryabko \cite{ryabko}]\rm Suppose \label{prop3}
\begin{equation}\label{eq13}
\lim_{i\rightarrow \infty}h(f_i^\infty)=h(f^\infty)\ .
\end{equation}
Then,
\begin{equation}\label{eq71}
\frac{1}{n}\log \frac{f^n(x_1,\cdots,x_n)}{g^n(x_1,\cdots,x_n)}\rightarrow 0
\end{equation}
 with probability 1, and
\begin{equation}\label{eq72}
\frac{1}{n}\int f^n(x_1,\cdots,x_n)dx_1\cdots dx_n
\log \frac{f^n(x_1,\cdots,x_n)}{g^n(x_1,\cdots,x_n)}\rightarrow 0
\end{equation}
as $n\rightarrow \infty$.
\end{prop}


\section{NonParametric Estimation for General Sources}\label{sec4}

Let $\{X_n\}_{n=1}^\infty$ be a stationary ergodic source expressed by measure  $\mu^\infty$
generating each $X_n$. 

Let $\{A_i\}_{i=0}^\infty$ be a sequence of finite partitions of $\mathbb R$
such that $A_{i+1}$ is a refinement of $A_i$, so that each $x\in A_i$
is a non-empty disjoint subset of $\mathbb R$.
For  each $i=0,1,\cdots$, since $P_i^\infty$ is stationary ergodic over the finite alphabet $A_i$, 
we can construct a universal coding $\varphi_i^n$ for $A_i^n$, and define
$$Q_i^n(a_1,\cdots,a_n):=
{2^{-|\varphi_i^n(a_1,\cdots,a_n)|}}
\ ,$$
for $a_1,\cdots,a_n\in A_i$. For measure $\eta^n: {\cal B}^n\rightarrow {\mathbb R}$ such that $\mu^n<<\eta^n$
and each $i=0,1,\cdots$, we define
\begin{eqnarray*}
&&\mu_i^n(D_1,\cdots,D_n)\\
&:=&\sum_{a_1,\cdots,a_n\in A_i}\frac{\eta^n(a_1\cap D_1,\cdots,a_n\cap D_n)}
{\eta^n(a_1,\cdots,a_n)}{P_i^n(a_1,\cdots,a_n)}
\end{eqnarray*}
and
\begin{eqnarray*}
&&\nu_i^n(D_1,\cdots,D_n)\\
&:=&\sum_{a_1,\cdots,a_n\in A_i}\frac{\eta^n(a_1\cap D_1,\cdots,a_n\cap D_n)}{\eta^n(a_1,\cdots,a_n)}{Q_i^n(a_1,\cdots,a_n)}
\end{eqnarray*}
for $(D_1,\cdots,D_n)\in {\cal B}^n$. 

Let $\{\omega_i\}_{i=0}^\infty$ be as in the previous section.
We  define the density estimation $\nu^n$ as follows: for $(D_1,\cdots,D_n)\in {\cal B}^n$, $i=0,1,\cdots$,
\begin{equation}\label{eq03}
\nu^n(D_1,\cdots,D_n):=\sum_{i=0}^\infty \omega_i \nu_i^n(D_1,\cdots,D_n)
\end{equation}
\begin{remark}\rm
When $\mu^n<<\eta^n=\lambda^n$, if we differentiate (\ref{eq03}) with $\lambda^n$, we obtain
$$\frac{d\nu^n}{d\lambda^n}(x_1,\cdots,x_n)=\sum_{i=0}^\infty \omega_i \frac{d\nu_i^n}{d\lambda^n}(x_1,\cdots,x_n)\ .$$
\end{remark}
If we put
$$g(x_1,\cdots,x_n):=\frac{d\nu^n}{d\lambda^n}(x_1,\cdots,x_n)$$
and
$$g_i(x_1,\cdots,x_n):=\frac{d\nu_i^n}{d\lambda^n}(x_1,\cdots,x_n)\ ,$$
then we see that (\ref{eq03}) becomes (\ref{eq02}) when $\mu^n<<\lambda^n$.

Notice that for any sequence of universal codes $\{\varphi_i^n\}_{i=0}^\infty$, 
we have $\nu^n(D^n)>0$ for all $D^n\in \{{\cal B}-\{\}\}^n$, so that $\mu^n << \nu^n$.

\begin{prop}[Barron \cite{barron}]\rm \label{prop6}
Let $(\Omega,{\cal F},\mu)$ and $\nu$ be the probability space and a $\sigma$-fnite measure.
Then,
\begin{equation}\label{eq127}
\frac{1}{n}\log \frac{d\mu^n}{d\nu^n}(x_1,\cdots,x_n)\rightarrow D(\mu||\nu):=\lim_{k\rightarrow \infty}\frac{1}{k}D(\mu^k||\nu^k)
\end{equation}
as $n\rightarrow \infty$ with probability one, where
\begin{equation}\label{eq128}
D(\mu^n||\nu^n):=\int d\mu^n\log \frac{d\mu^n}{d\nu^n}\geq 0
\end{equation}
if $\nu^n(\Omega)\leq 1$.
\end{prop}

From Proposition \ref{prop6}, since $\eta^n(\Omega)\leq 1$, we have 
$$\frac{1}{n}\log \frac{d\mu^n}{d\eta^n}(x_1,\cdots,x_n)\rightarrow D(\mu||\eta)\geq 0$$
and
$$\frac{1}{n}\log \frac{d\mu_i^n}{d\eta^n}(x_1,\cdots,x_n)\rightarrow D(\mu_i||\eta)\geq 0$$
as $n\rightarrow \infty$ with probability one.

\begin{teiri}\rm Suppose
\begin{equation}\label{eq87}
\lim_{i\rightarrow \infty}D(\mu_i||\eta)=D(\mu_||\eta).
\end{equation}
Then,
\begin{equation}\label{eq89}
\frac{1}{n}\log \frac{d\mu^n}{d\nu^n}(x_1,\cdots,x_n)\rightarrow 0
\end{equation}
and
\begin{equation}\label{eq88}
D(\mu||\nu)=0 \ .
\end{equation}
\end{teiri}
Proof. For any integer $i$, we have
$$\nu^n(D_1,\cdots,D_n)\geq \omega_i\nu_i^n(D_1,\cdots,D_n)$$
for any $(D_1,\cdots,D_n)\in {\cal B}^n$.
which is equivalent to
$$\frac{d\mu^n}{d\nu^n}(x_1,\cdots,x_n)\leq \frac{1}{\omega_i}\frac{d\mu^n}{d\nu_i^n}(x_1,\cdots,x_n)$$
with probability 1.
Hence,
\begin{eqnarray}\label{eq10}
&&\frac{1}{n}\log \frac{d\mu^n}{d\nu^n}(x_1,\cdots,x_n)\nonumber\\
&\leq&
- \frac{1}{n}\log \omega_i 
+
 \frac{1}{n}\log \frac{d\mu_i^n}{d\nu^n_{i}}(x_1,\cdots,x_n)\nonumber\\
 &&+
 \frac{1}{n}\log \frac{d\mu^n}{d\mu^n_{i}}(x_1,\cdots,x_n)
\end{eqnarray}
with probability 1 for each $i=0,1,\cdots$.
The first term converges to zero.
Since $\mu^\infty$ is stationary ergodic, so is $\mu_i^\infty$ for each $i=0,1,\cdots$. Thus, from Proposition \ref{prop1},

$$\frac{1}{n}\log \frac{P_i^n(a_1,\cdots,a_n)}{Q_i^n(a_1,\cdots,a_n)}\rightarrow 0$$ with probability 1
as $n\rightarrow \infty$, so that we may write
$$P_i^n(a_1,\cdots,a_n)=Q_i^n(a_1,\cdots,a_n)f_i(a_1,\cdots,a_n)$$
with $f_i(a_1,\cdots,a_n)^{1/n}\rightarrow 1$. If we define 
$$K_{i,n}:=\max_{a_1,\cdots,a_n\in A_i}f_i(a_1,\cdots,a_n)\ ,$$ then
\begin{eqnarray*}
&&\frac{1}{n}\log \frac{\mu_i(D_1,\cdots,D_n)}{\nu_i(D_1,\cdots,D_n)}\\
&=&\frac{1}{n}\log \frac{
\sum_{a_1,\cdots,a_n\in A_i}\frac{\eta^n(a_1\cap D_1,\cdots,a_n\cap D_n)}{\eta^n(a_1,\cdots,a_n)}{P_i^n(a_1,\cdots,a_n)}}
{
\sum_{a_1,\cdots,a_n\in A_i}\frac{\eta^n(a_1\cap D_1,\cdots,a_n\cap D_n)}{\eta^n(a_1,\cdots,a_n)}{Q_i^n(a_1,\cdots,a_n)}}\\
&\leq&\frac{1}{n}\log K_{i,n}\rightarrow 0\ ,
\end{eqnarray*}
which means that the second term in (\ref{eq10}) also converges to zero as $n\rightarrow \infty$.

From (\ref{eq127})(\ref{eq87})(\ref{eq10}), we have the first statement (\ref{eq89}).
The second equation (\ref{eq88}) immediate from Proposition \ref{prop6}.

\section{Examples}

The condition (\ref{eq87}) in Theorem 1 exactly specifies what $\{A_i\}_{i=0}^\infty$ should satisfy.

\begin{rei}[Finite Sources]\rm
Since
$$\mu^n(a_1,\cdots,a_n)=P^n(a_1,\dots,a_n)$$
and 
$$\nu^n(a_1,\cdots,a_n)=Q^n(a_1,\dots,a_n)$$
for $a_1,\cdots,a_n\in A$, 
(\ref{eq89})(\ref{eq88}) extend 
(\ref{eq301})(\ref{eq302}), respectively, and condition
(\ref{eq87}) is always met. 
\end{rei}
\begin{rei}[Continuous Sources]\rm
The equations (\ref{eq87})(\ref{eq89})(\ref{eq88}) extend 
(\ref{eq13})(\ref{eq71})(\ref{eq72}) in Proposition \ref{prop3}.
In fact, when $\mu^n <<\lambda^n$ and $\mu_i^n << \lambda^n$, we have
$$\frac{d\mu^n}{d\lambda^n}(x_1,\cdots,x_n)=f^n(x_1,\cdots,x_n)$$
and
$$\frac{d\mu_i^n}{d\lambda^n}(x_1,\cdots,x_n)=f^n_i(x_1,\cdots,x_n)\ .$$
For example, (\ref{eq87}) implies
$$
\lim_{i\rightarrow \infty}\lim_{n\rightarrow \infty}\frac{1}{n}\log \frac{\frac{d\mu^n}{d\lambda^n}(x_1,\cdots,x_n)}{\frac{d\mu_i^n}{d\lambda^n}(x_1,\cdots,x_n)}=0\ ,
$$
which further from (\ref{eq201})(\ref{eq202}) implies (\ref{eq13}).
In particular, if the source is i.i.d. and that the density function 
$$f^n(x_1,\cdots,x_n)=\prod_{j=1}^n f(x_j)$$
is both continuous and positive only when $a\leq x_j\leq b$, $j=1,\cdots,n$, then for 
$A_i:=\{c_0,\cdots,c_{2^{i+1}-1}\}$ with $c_k=[a+k\cdot 2^{-i-1}(b-a),a+(k+1)\cdot 2^{-i-1}(b-a))$,
$k=0,1,\cdots,2^{i+1}-1$, there exists $\{M_i\}$ such that 
$$\sup_k\sup_{x,y\in c_k}|f(x)-f(y)|<M_i\rightarrow 0$$
as $i\rightarrow \infty$. Then, for each $x\in A$ such that $f_i(x)>0$,
$$\log \frac{f(x)}{f_i(x)}<\log \frac{f_i(x)+M_i}{f_i(x)}\rightarrow 0$$
and
$$\log \frac{f_i(x)}{f(x)}<\log \frac{f(x)+M_i}{f(x)}\rightarrow 0$$
as $i\rightarrow \infty$, where the mean value theorem:
 if $a\leq r<s\leq b$, there exists $r\leq x_0\leq s$ such that
 $$\frac{1}{s-r}\int_r^sf(x)dx=f(x_0)$$
 has been applied. Hence, we have
\begin{eqnarray*}
&&|\frac{1}{n}\log \frac{f^n(x_1,\cdots,x_n)}{f_i^n(x_1,\cdots,x_n)}|\\
&\leq &\frac{1}{n}\sum_{j=1}^n\max \{\log \frac{f_i(x_j)+M_i}{f_i(x_j)},\log \frac{f(x_j)+M_i}{f(x_j)}\}\\
&\leq &\frac{1}{n}\sum_{j=1}^n \log \frac{(f(x_j)+M_i)(f_i(x_j)+M_i)}{f(x_j)f_i(x_j)}
\end{eqnarray*}
which almost surely converges to a constant $L_i$ as $n\rightarrow \infty$ such that $\lim_{i\rightarrow \infty}L_i=0$.
Thus, condition (\ref{eq87}) is obtained. 
\end{rei}
\begin{rei}[Countable Sources]
Suppose the source is i.i.d. and that $X(\Omega)={\mathbb N}=\{1,2,\cdots\}$.
If we choose $\displaystyle \eta(X=k):=\frac{1}{k}-\frac{1}{k+1}$ and $A_i:=\{\{1\},\cdots,\{i\},{\mathbb N}-\{1,\cdots,i\}\}$,
then 
$$\mu_i(X=k)=
\left\{
\begin{array}{ll}
\mu(X=k),& k\leq i\\
\displaystyle \{\sum_{l=i+1}^\infty\mu(X=l)\} \frac{i+1}{k(k+1)},& k\geq i+1\ ,
\end{array}
\right.
$$
so that $\mu_i(X=k)\rightarrow \mu(X=k)$ as $i\rightarrow \infty$ for all $k\in {\mathbb N}$.
Thus,
$\displaystyle \frac{1}{n}\frac{d\mu^n}{d\mu_i^n}(x_1,\cdots,x_n)$ almost surely converges to a constant $L_i$ such that $L_i\rightarrow 0$ as $i\rightarrow \infty$.
\end{rei}

\section{On-Line Prediction}

From Lemma 1, and since $\mu^j<<\mu^{j-1}$ and $\nu^j <<\nu^{j-1}$,
there exist conditional measures
$\mu^{j|j-1}(D_j|x^{j-1})$ and $\nu^{j|j-1}(D_j|x^{j-1})$ for $D_j\in {\cal B}$ and $x^{j-1}\in {\mathbb R}^{j-1}.$
The conditional measure is used for online prediction.

\begin{teiri}\rm
Let $r:{\mathbb R}\rightarrow {\mathbb R}$ be a bounded measurable function, then under (\ref{eq87}),
\begin{eqnarray*}
&&\lim_{n\rightarrow \infty}\frac{1}{n}E\sum_{j=0}^{n-1}
\{\int r(x)d\mu^{j|j-1}(x|x_1,\cdots,x_{j-1})\\
&&-
\int r(x)d\nu^{j|j-1}(x|x_1,\cdots,x_{j-1})
\}^2
=0
\end{eqnarray*}
and
\begin{eqnarray*}
&&\lim_{n\rightarrow \infty}\frac{1}{n}E\sum_{j=0}^{n-1}|
\int r(x)d\mu^{j|j-1}(x|x_1,\cdots,x_{j-1})\\
&&-
\int r(x)d\nu^{j|j-1}(x|x_1,\cdots,x_{j-1})|
=0\ .
\end{eqnarray*}
\end{teiri}
Proof: Let $b$ be such that $|r(x)|<b$, $x\in {\mathbb R}$. Then, for each $j=1,\cdots,n$, we have
\begin{eqnarray*}
&&\{\int r(x)d\mu^{j|j-1}(x|x_1,\cdots,x_{j-1})\\
&&-
\int r(x)d\nu^{j|j-1}(x|x_1,\cdots,x_{j-1})
\}^2\\
&\leq&b^2\{\int d\mu^{j|j-1}(x|x_1,\cdots,x_{j-1})\\
&&-
\int d\nu^{j|j-1}(x|x_1,\cdots,x_{j-1})
\}^2\\
&\leq&{b^2}\{\sup_A|\int_A d\mu^{j|j-1}(x|x_1,\cdots,x_{j-1})\\
&&-
\int_A d\nu^{j|j-1}(x|x_1,\cdots,x_{j-1})|
\}^2\\
&\leq&\frac{2b^2}{\log e}\int d\mu^{j|j-1}(x|x_1,\cdots,x_{j-1})\\
&&\cdot \log \frac{d\mu^{j|j-1}}{d\nu^{j|j-1}}(x|x_1,\cdots,x_{j-1})\ ,
\end{eqnarray*}
where Lemma \ref{L7} below
has been applied for the last inequality.
From Theorem 1, we obtain the final result:
\begin{eqnarray*}
&&\frac{1}{n}[E\sum_{j=1}^{n}|\int r(x)d\mu^{j|j-1}(x|x_1,\cdots,x_{j-1})\\
&&-
\int r(x)d\nu^{j|j-1}(x|x_1,\cdots,x_{j-1})|]^2\\
&\leq &\frac{1}{n}E\sum_{j=1}^{n}\{\int r(x)d\mu^{j|j-1}(x|x_1,\cdots,x_{j-1})\\
&&-
\int r(x)d\nu^{j|j-1}(x|x_1,\cdots,x_{j-1})
\}^2\\
&\leq&\frac{2b^2}{\log e}\cdot \frac{1}{n} \int d\mu^{n}(x_1,\cdots,x_n)\log \frac{d\mu^{n}}{d\nu^{n}}(x_1,\cdots,x_n)\\
&\rightarrow& 0
\end{eqnarray*}
\begin{hodai}[Pinsker's inequality \cite{gallager}]\rm \label{L7}
$$\sup_{A \in {\cal F}}|\mu(A)-\nu(A)|
\leq \sqrt{\frac{2}{\log e} D(\mu||\nu)}$$
\end{hodai}

\begin{rei}[Finite Sources]\rm \label{prop2}
Let $\{X_n\}_{n=1}^\infty$ be a stationary ergodic source  $P^\infty$ with each 
$X_n$ in a finite set $A$.
Let $r: A\rightarrow {\mathbb R}$ be a bounded  integrable function, then
\begin{eqnarray*}
&&\lim_{n\rightarrow \infty}\frac{1}{n}E\sum_{j=0}^{n-1}
\{\sum_{x\in A} r(x)P^{j|j-1}(x|x_1,\cdots,x_{j-1})\\
&&-
\sum_{x\in A} r(x)Q^{j|j-1}(x|x_1,\cdots,x_{j-1})
\}^2
=0
\end{eqnarray*}
and
\begin{eqnarray*}
&&\lim_{n\rightarrow \infty}\frac{1}{n}E\sum_{j=0}^{n-1}|
\sum_{x\in A} r(x)P^{j|j-1}(x|x_1,\cdots,x_{j-1})\\
&&-
\sum_{x\in A} r(x)Q^{j|j-1}(x|x_1,\cdots,x_{j-1})|
=0\ .
\end{eqnarray*}
where
$$P^{j|j-1}(x_j|x_1,\cdots,x_{j-1}):=\frac{P^{j}(x_1,\cdots,x_{j-1},x_{j})}{P^{j-1}(x_1,\cdots,x_{j-1})}$$
and
$$Q^{j|j-1}(x_j|x_1,\cdots,x_{j-1}):=\frac{Q^{j}(x_1,\cdots,x_{j-1},x_{j})}{Q^{j-1}(x_1,\cdots,x_{j-1})}\ .$$
\end{rei}

\section{Concluding Remarks}

We proposed a learning algorithm for nonparametric estimation and
on-line prediction for general stationary ergodic sources.
However, we need to know a priori what $\{A_i\}$ satisfies (\ref{eq87}).
In this sense, rough knowledge about the true $mu$ is required.
Also, although the sequence $\{\omega_i\}$ is infinite to achieve the desired properties such as
(\ref{eq88}) (\ref{eq89}), however, we can only use finite $\{\omega_i\}$ in reality.

Future topic includes a way to choose the sequences $\{A_i\}$ and $\{\omega_i\}$
for getting a better solution for finite $n$ by utilizing the a prior knowledge of the probability measure.

\section*{Appendix: Measure theory}
Let $\cal F$ be a $\sigma$-field of entire set $\Omega$, and $\mu,\nu$ $\sigma$-finite measures \cite{billingsley} on $\cal F$,
which means that there exist $\{A_k\}$ and $\{B_k\}$ such that $\Omega=\cup_k A_k=\cup_k B_k$ and $\mu(A_k)<\infty, \nu(B_k)<\infty$, $k=1,2,\cdots$.
We say $g:\Omega \rightarrow {\mathbb R}$ is ${\cal F}$-masurable if $\{\omega\in \Omega|g(\omega)\in D\}\in {\cal F}$
for each $D\in {\cal B}$, where $\cal B$ is the Borel $\sigma$-field.
We write $\mu<<\nu$ when $\nu(A)=0 \Longrightarrow \mu(A)=0$ for $A\in {\cal F}$, and define
the integral for ${\cal F}$-measurable $g: \Omega\rightarrow {\mathbb R}$
over $A\in {\cal F}$ with respect to $\nu$
 by
$$\int_A g(\omega)d\nu(\omega):=\sup_{\{A_i\}}\sum_{i} \inf_{\omega\in A_i} g(\omega)\nu(A_i)\ ,$$
where $\{A_i\}$ ranges over the whole partitions of $A$.

\begin{hodai}[Radon-Nykodim \cite{billingsley}]\rm \label{L0}
Suppose $\mu<<\nu$ and that they are both $\sigma$-finite. There exists a $\cal F$-measurable $\displaystyle \frac{d\mu}{d\nu}:=g:{\Omega}\rightarrow {\mathbb R}$ such that
$$\mu(A)=\int_A g(\omega)d\nu(\omega)$$
for $A\in {\cal F}$. In particular,
$$\mu<<\nu <<\lambda \Longrightarrow \frac{d\mu}{d\nu}\cdot \frac{d\nu}{d\lambda} =\frac{d\mu}{d\nu}\ .$$
\end{hodai}

Let $(\Omega,{\cal F},\mu)$ be a probability space and $\nu$ a measure such that $\nu(\Omega)\leq 1$ and $\mu<<\nu$.
\begin{teigi}[Kullback-Leibler information \cite{kullback}]\rm
Suppose\footnote{The base two logarithm is assumed throughout the paper.} $\mu<<\nu$.
$$D(\mu||\nu):=\int_{\Omega}d\mu(\omega)\log \frac{d\mu}{d\nu}(\omega)\ .$$
\end{teigi}
Any $\cal F$-measurable map $X:\Omega\rightarrow {\mathbb R}$ is said a random variable.


\end{document}